# Fractal and multifractal analysis of complex networks: Estonian network of payments

Stephanie Rendón de la Torre[*], Jaan Kalda[*], Robert Kitt[*,1], Jüri Engelbrecht[*]

[*]School of Science, Department of Cybernetics, Tallinn University of Technology, Akadeemia tee 21, 12618, Tallinn, ESTONIA
[1]Swedbank AS, Liivalaia 12, 15038, Tallinn, ESTONIA

______________________________________________________________________

**Abstract**

Complex networks have gained much attention from different areas of knowledge in recent years. Particularly, the structures and dynamics of such systems have attracted considerable interest. Complex networks may have characteristics of multifractality. In this study, we analyze fractal and multifractal properties of a novel network: the large scale economic network of payments of Estonia, where companies are represented by nodes and the payments done between companies are represented by links. We present a fractal scaling analysis and examine the multifractal behavior of this network by using a sandbox algorithm. Our results indicate the existence of multifractality in this network and consequently, the existence of multifractality in the Estonian economy. To the best of our knowledge, this is the first study that analyzes multifractality of a complex network of payments.

*Corresponding author.
E-mail Address: stretomx@gmail.com (S. Rendón de la Torre)

## 1. Introduction

In recent years, complex networks have been studied extensively and have attracted much attention from researchers belonging to different fields of knowledge and science. Complex networks theory is developing at a fast pace and has already made significant progress towards designing the framework for unraveling the organizing principles that govern complex networks and their evolution. In fact, several topological characteristics and a variety of dynamical aspects of complex networks have been the center of extensive research and studies in the last years.

A fractal is a quantity or a fragmented geometric object which can be split into parts, each of which is a reduced-size copy of the whole and has the same statistical character as the whole. A fractal displays self-similarity on all scales. Fractals have infinitely complex patterns that are self-similar across different scales; these objects do not need to display exactly the same structures at all scales, but the same "type" of structures must appear on all scales. Fractals can be created by repeating simple recursive processes. Another definition of a fractal states that is a set whose Hausdorff-Besicovitch dimension strictly exceeds its topological dimension. The "fractal geometry" of nature was first labeled as a term by Benoit Mandelbrot in the late 60's, and after that, the fractal approach has been widely used to gain insight into the fundamental scaling of numerous complex structures. Fractal analysis helps to distinguish global features of complex networks, such as the fractal dimension. However, the fractal formalism is insufficient to characterize the complexity of many real networks which cannot simply be described by a single fractal dimension. Furuya and Yakubo [1] demonstrated analytically and numerically that fractal scale-free networks may have multifractal structures in which the fractal dimension is not sufficient to describe the multiple fractal patterns of such networks, therefore, multifractal analysis rises as a natural step after fractal analysis.

Multifractal structures are abundant in social systems and are plentiful in a variety of physical phenomena. Multifractal analysis is a systematic approach and a generalization of fractal analysis that is useful when describing spatial heterogeneity of fractal patterns [2]. It has proven to be a useful tool for studying turbulence phenomena [3-4], time series analysis [5-6], economic and financial modeling [7], medical pattern recognition [8], biological and geophysical systems [9-18].

Fractal and multifractal analysis can help to uncover the structure of all kinds of systems in order to have a better understanding of such systems. In particular, both approaches have many different interesting applications in economy. An interesting line of research is related with the relevance and applicability of fractal and multifractal analysis in social and economic topics. Inaoka *et al.* [19] showed that the study of the structure of a banking network provides useful insight from practical points of view. By knowing and understanding the network structure and characteristics of banking networks (in terms of transactions and their patterns), a systemic contagion could potentially be prevented. In their study, these authors showed that the network of financial transactions of Japanese financial institutions has a fractal structure. Regarding social studies, Lu *et al*. [20] showed the importance of road patterns for urban transportation capacity based on fractal analysis of such network. In this study, the authors were able to link the fractal measurement with city mass measurements.

Another direction of these studies has tilted towards the development of multifractal models for financial networks [21]. A few recent studies have focused on the analysis of the changes of multifractal spectra across time to assess changes in economy during crisis periods [22-23]. Some other studies have focused on gathering empirical evidence of the common multifractal signature in economic, biological and physical systems [24].

In the last years, numerous algorithms for calculating the fractal dimension and studying self-similar properties of complex networks have been developed and tested extensively [25-29]. Song *et al.* [30] developed a method for calculating the fractal dimension of a complex network by using a box-covering algorithm and identified self-similarity as a property of complex networks [31]. Moreover, a myriad of algorithms and studies on networks' multifractal analysis have been proposed and developed lately [32-37].

The main objective of this study is to analyze the fractality and multifractality of a novel and unstudied network: the large scale Estonian network of payments. We present a study that contributes to the field of complex networks (particularly to economic complex networks studies) by adding empirical evidence in favor of fractalism and multifractalism with a new case of study. The study is done thanks to the application of known network methods. Also, the goal is to expand the knowledge of the structure of this network of payments by analyzing its fractal and multifractal structures anticipating that this analysis could be useful in the future for further studies. Multifractal analysis of a payment network could be the starting point for developing economical and financial future studies related with: opportune detection of key factors driving the multifractal spectrum changes across time, money-flows forecasting and risk-pattern recognition (during turbulent financial times, for example). To the extent of our knowledge, this is the first study that examines multifractality of a complex network of payments.

We present a fractal scaling analysis by calculating the fractal dimension of our network and its skeleton. Then, we use a sandbox algorithm for complex networks [33] to calculate the spectrum of the generalized fractal dimensions $D(q)$ and mass exponents $\tau(q)$ in order to study the multifractal behavior of the network. Section 1 presents an introduction and an overview of literature related with fractal and multifractal network studies. Section 2 is devoted to a detailed description of the data set and the methods used. Section 3 presents our main results and Section 4 concludes with a discussion of the results.

**2. Data and methodology**

In this section we describe the nature and the scope of our data set. Then, we introduce the box-covering algorithm used to calculate the fractal dimension of our network and its skeleton. To conclude, we introduce the sandbox algorithm used for multifractal analysis in this study.

**2.1 Data**

To create the network of payments we used payment transactions data from Swedbank. At present, Swedbank is one of the leading banks in the Nordic and Baltic regions of Europe and operates in the following countries: Latvia, Lithuania, Estonia and Sweden. The data and all the information related with the identities of the nodes are very sensitive and will remain confidential. Our data set is unique and very interesting because ~80% of Estonia's electronic bank transactions are executed through Swedbank's system of payments, hence, this data set reproduces fairly well the trends of money of the whole Estonian economy. Our Estonian network of payments focuses exclusively on domestic payments transferred electronically from customer to customer (company-to-company) during the year 2014. There are 16,613 nodes, 2,617,478 payments and 43,375 undirected links in the selected data set.

A network is a set of nodes connected by links. In this study, the nodes represent companies and the links represent the payments between the companies. We mapped a symmetric payments adjacency matrix $A_{NxN}$ where $N$ is the total number of nodes in the network. The payments adjacency matrix $A_{NxN}$ represents the whole image of the network. For simplicity, we considered an undirected graph approach where two nodes have a link if they share one or more payments, then each element represents a link if there is a transaction between companies $i$ and $j$ as follows: $a_{ij}^u = a_{ji}^u$ and $a_{ij}^u = 1$; otherwise, $a_{ij}^u = 0$ if there is no transaction between companies $i$ and $j$.

**2.2 Fractal scaling analysis**

Fractal analysis assists on the calculation and the understanding of the fractal dimension of complex networks. In general, fractal analysis consists of several methods to measure complexity by using the fractal dimension and other fractal characteristics. According to Song *et al.* [31] complex networks may have self-similar structures. According to these authors, the box-counting algorithm is an appropriate method to examine global properties of complex networks. The fundamental relation of fractal scaling is based on the box-covering method which counts the total number of boxes that are needed to cover a network with boxes of certain size. The box-covering method is analogous to the box-counting method widely used in fractal geometry and is a basic tool to measure the fractal dimension of fractal objects embedded in Euclidean space [38]. However, the Euclidean metric is not well defined for networks, thus we use the adaptation for networks developed by Wang *et al.* [35] of the random sequential box-covering algorithm created by Kim *et al.* [39], to determine the fractal dimension of our network and its skeleton. The aforementioned method contains a random process for selecting the position of the center of each box. We let $N_B(r_B)$ be the minimum number of boxes needed to tile the whole network, where the lateral size of the boxes is the measure of radius $r_B$ as follows:

$$N_B(r_B) \sim r_B^{-d_B}, \tag{1}$$

where $d_B$ is the fractal dimension. If we measure the number $N_B$ for different box sizes, then it is possible to obtain the fractal dimension $d_B$ by obtaining the power law fitting of the distribution. The algorithm selects a random node at each step, and this node is the seed that will be the center of a box. Then, we search the network by distance $r_B$ from the seed node and cover all the nodes that are located within that distance, but only if they have not been covered yet. Then, we assign the newly covered nodes to the new box; if there are no more newly covered nodes then the box is removed. This process is repeated until all the nodes of the network belong to boxes.

Before using the algorithm, we calculate the skeleton of the network. The concept of skeleton was first introduced by Kim *et al.* [40]. The skeleton is a particular type of spanning tree based on the link betweenness centrality (a simplified quantity to measure the traffic of networks) that is entrenched beneath the original network. The skeleton delivers a shell for the fractality of the network and is formed by links with the highest betweenness centralities. Only the links that do not form loops are included. The remaining links from the original network which are not included in the skeleton are local shortcuts that contribute to loop formation, meaning that the distance between any two nodes in the original network may increase in the skeleton. A fractal network has a fractal skeleton beneath it which is distressed by these local shortcuts, but preserving its fractality. For a scale-free network, its skeleton also follows a power-law degree distribution where its degree exponent might differ slightly from that of the original network. When studying the origin of fractality in networks, actually the skeleton is more useful than the original network itself due to its unsophisticated tree structure [41]. In general, the skeleton preferentially collects the sections of the network where betweenness is high, and this preserves the structure and simplifies its complexity. Consequently, by looking at the properties of the skeleton it is easier to appreciate the topological organization of the original network.

To calculate the skeleton of a complex network, the link betweenness of all the links in the network has to be calculated. The betweenness centrality of a network (for a link or a node), is defined as follows:

$$b_i = \sum_{j,k \in N, j \neq k} \frac{n_{jk}(i)}{n_{jk}} \qquad (2)$$

where $N$ is the total number of nodes, $n_{jk}$ is the total number of shortest-paths between nodes $j$ and $k$, $n_{jk}(i)$ is the total number of shortest-paths linking nodes $j$ and $k$ that passes through the node $i$.

In order to perform the fractal scaling analysis, we used Dijkstra's algorithm [42]; then we used the box-covering algorithm to calculate the fractal dimension of the network and the skeleton to compare both values.

**2.3 Sandbox algorithm for multifractal analysis of complex networks**

Scale-free networks are commonly observed in a wide array of different contexts of nature and society. Previous studies [43-44] have shown that in scale-free networks, independently of the system and the identity of their components, the probability $P(k)$ that a node in the network interacts with $k$ other links decays as a power-law, following that $P(k) \sim k^{-\gamma}, k \neq 0$; this points to a tendency for large networks to self-organize into a scale-free state. We found scale-free properties characterized by power-law degree distributions in our previous study on the Estonian network of payments [45] (see Tables 1 and 2 for details of the main features and statistics of this network).

In general, multifractality is expected to appear in scale-free networks due to the fluctuations that occur in the density of local nodes. Multifractal analysis requires taking into account a physical measure, like the number of nodes within a box of specific size in order to analyze how the distribution of such number of nodes scales in a network as the size of the box grows. Tél *et al.* [46] introduced a sandbox algorithm based on the fixed-size box-counting algorithm [47] which was used and adapted for multifractal analysis of complex networks by Liu *et al.* [33]. In order to determine the multifractal dimensions of our complex network, we chose this adapted sandbox algorithm because it is precise, efficient and practical. Moreover, a study by Song *et al.* [2] has shown that this algorithm gives better results when it is used in unweighted networks, and this is our case.

The fixed-size box-counting algorithm is one of the most known and efficient algorithms for multifractal analysis. For a given probability measure $0 \leq \mu \leq 1$ in a metric space $\Omega$ with a support set $E$, we consider the following partition sum:

$$Z_\varepsilon(q) = \sum_{\mu(B) \neq 0} [\mu(B)]^q, \qquad (3)$$

where the parameter $q \in R$, and it describes the moment of the measure. The sum runs over all different non-overlapping (or non-empty) boxes $B$ of a given size ε that covers the support set $E$. From this definition, it is easy to obtain $Z_\varepsilon(q) \geq 0$ and $Z_\varepsilon(0) = 1$. The function of the mass exponents $\tau(q)$ of the measure $\mu$ is defined by:

$$\tau(q) = \lim_{\varepsilon \to 0} \left( \frac{\ln Z_\varepsilon(q)}{\ln \varepsilon} \right). \qquad (4)$$

Then, the generalized fractal dimensions $D(q)$ of the measure $\mu$ are defined as follows:

$$D(q) = \frac{\tau(q)}{q-1}, q \neq 1, \tag{5}$$

and

$$D(1) = \lim_{\varepsilon \to 0} \frac{Z_{(1,\varepsilon)}}{\ln \varepsilon}, q = 1, \tag{6}$$

where

$$Z_{1,\varepsilon} = \sum_{\mu(B) \neq 0} \mu(B) \ln \mu(B). \tag{7}$$

The generalized fractal dimensions $D(q)$ can be estimated with linear regression of $[\ln Z_\varepsilon(q)]/[q-1]$ against $\ln \varepsilon$ for $q \neq 1$, and similarly a linear regression of $Z_{1,\varepsilon}$ against $\ln \varepsilon$ for $q = 1$. $D(0)$ is the fractal dimension or the box-counting dimension of the support set $E$ of the measure $\mu$, $D(1)$ is the information dimension and $D(2)$ is the correlation dimension.

For a complex network, a box of size $B$ can be defined in terms of the distance $l_B$, which corresponds to the number of links in the shortest-path between two nodes. This means that every node is less than $l_B$ links away from another node in the same box. The measure $\mu$ of each box is defined as the ratio of the number of nodes that are covered by the box and the total number of nodes in the whole network.

Multifractality of a complex network can be determined by the shape of $\tau(q)$ or $D(q)$ curves. If $\tau(q)$ is a straight line or $D(q)$ is a constant, then the network is monofractal; similarly if $D(q)$ or $\tau(q)$ have convex shapes, then the network is multifractal. A multifractal structure can be identified by the following signs [48]: a) multiple slopes of $\tau(q)$ vs $q$; b) non-constant $D(q)$ vs $(q)$ values and c) $f(\alpha)$ vs $\alpha$ value covers a broad range (not accumulated at nearby non-integer values of $\alpha$).

First, we calculate the shortest-path distance between any two nodes in the network and map the shortest-path adjacency matrix $B_{NxN}$ using the payments adjacency matrix $A_{NxN}$. We use the shortest-path adjacency matrix $B_{NxN}$ as input for multifractal analysis. The central idea of the sandbox algorithm is simply to select a node of the network in a random fashion as the center of a sandbox and then count the number of nodes that are inside the sandbox. Initially, none of the nodes has been chosen as a center of a box or as a seed. We set the radius $r$ of the sandbox which will be used to cover the nodes in the range $r \in [1, d]$, where $d$ (diameter) is the longest distance between nodes in the network and radii $r$ are integer numbers. We ensure that the nodes are chosen randomly as center nodes by reordering the nodes randomly in the whole network. Depending on the size $N$ of the network, we choose $T$ nodes in random order as centers of $T$ sandboxes; then we find all the neighboring nodes within radius $r$ from the center of each box. We count the number of nodes contained in each sandbox of radius $r$, and denote it by $S(r)$. We calculate the statistical averages $\langle [S(r)^{q-1}] \rangle$ of $[S(r)^{q-1}]$ over all the sandboxes $T$ of radius $r$. The previous steps will be repeated for each of the different values of radius $r$ to obtain the statistical average $\langle [S(r)^{q-1}] \rangle$ and then use it for calculating linear regression.

The generalized fractal dimensions $D(q)$ of the measure $\mu$ are defined by

$$D(q) = \lim_{r \to 0} \frac{\ln\langle [S(r)/S(0)]^{q-1}\rangle}{\ln(r/d)} \frac{1}{q-1}, q \in R, \quad (8)$$

or rewritten as

$$\ln(\langle [S(r)]^{q-1}\rangle) \propto D(q)(q-1)\ln(r/d) + (q-1)\ln(S_0), \quad (9)$$

where $S(0)$ is the size of the network and the brackets mean taking statistical average over the random selection of the sandbox centers.

We run the linear regression of $\ln(\langle [S(r)]^{q-1}\rangle)$ against $(q-1)\ln(r/d)$ to obtain the generalized fractal dimensions and similarly, calculate the linear regression of $\ln(\langle [S(r)]^{q-1}\rangle)$ against $\ln(r/d)$ to obtain the mass exponents $\tau(q)$. From the shapes of the generalized fractal dimensions curves, we can conclude if multifractality exists or not in our network.

## 3. Results

### 3.1 Fractal scaling analysis

The general characteristics and statistics of the Estonian network of payments are listed in Tables 1 and 2.

Table 1 Network's characteristics

| | |
|---|---|
| Companies analyzed | 16,613 |
| Total number of payments analyzed | 2,617,478 |
| Value of transactions | 3,803,462,026 * |
| Average value of transaction per customer | 87,600 * |
| Max value of transaction | 121,533 * |
| Min value of transaction (aggregated in whole year) | 1,000 * |
| Average volume of transaction per company | 60 |
| Max volume of transaction per company | 24,859 |
| Min volume of transaction per company (aggregated in whole year) | 20 |

* All money amounts are expressed in monetary units and not in currencies in order to protect the confidentiality of the data set. The purpose of showing monetary units is to provide a notion of the proportion of quantities and not to show exact amounts of money.

Table 2 Summary of Statistics

| Statistic | Value | Components | # of nodes |
|---|---|---|---|
| Undirected Links | 43,375 | GCC | 15,434 |
| $<k>$ | 20 | DC | 1,179 |
| $\gamma^o$ | 2.39 | GSCC | 3,987 |
| $\gamma^i$ | 2.49 | GOUT | 6,054 |
| $\gamma$ | 2.45 | GIN | 6,172 |
| $<C>$ | 0.183 | Tendrils | 400 |
| $<l>$ | 7.1 | Cutpoints | 1,401 |
| $T$ | 0.13 | Bi-component | 4,404 |
| $D$ | 29 | $k$-core | 1,081 |
| $<\sigma>$ (nodes) | 110 | | |
| $<\sigma>$ (links) | 40 | | |

N = number of nodes. $<k>$ = average degree. $\gamma^o$ = scaling exponent of the out-degree empirical distribution. $\gamma^i$ = scaling exponent of the in-degree empirical distribution. $\gamma$ = scaling exponent of the connectivity degree distribution. $<C>$ = average clustering coefficient. $<l>$ = average shortest path length. $T$ = connectivity %. $D$ = Diameter. $<\sigma>$ = average betweenness. GCC = Giant Connected Component. DC = Disconnected Component. GSCC = Giant Strong Connected Component. GOUT = Giant Out Component. GIN = Giant In Component.

We present a fractal scaling analysis by using the box-counting algorithm expressed in Equation 1. We calculated the fractal dimension of our network and its skeleton (see Figures 1 and 2), where the fractal dimension is the absolute value of the slope of the linear fit. Figure 1 depicts a visualization of the graph representation of the skeleton of our network. The box-covering method yields a fractal dimension $d_{Bs}= 2.32 \pm 0.07$ for the skeleton network and for the original network the fractal dimension is $d_{Bo}= 2.39 \pm 0.05$.

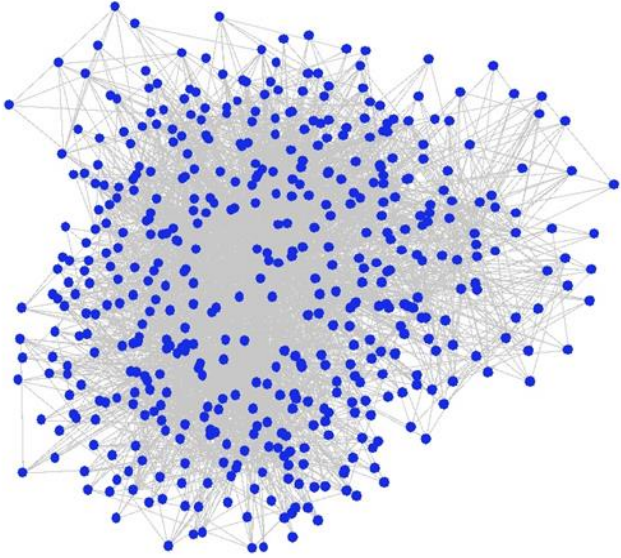

Figure 1. Visual graph representation of the skeleton of the Estonian network of payments.

The comparison of the fractal scaling in our network and its skeleton structure revealed its own patterns according to the fractality of the network. Figure 2 depicts the fractal scaling representation of our network. As seen in Figure 2, the respective number of boxes needed to cover both networks is similar but not identical: more boxes were needed for covering the skeleton. The largest distance between any two nodes in the network of payments is 29, while the largest distance between any two nodes in the skeleton network is 34.

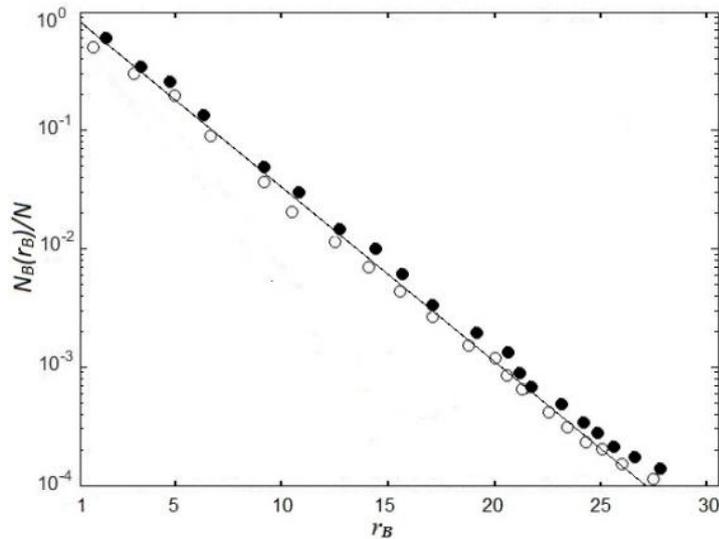

Figure 2. Fractal scaling representation of our network. The original network (o) and the skeleton network (●). The straight line is included for guidance and has a slope of 2.3. The analysis includes only the giant connected cluster of the network.

## 3.2 Multifractal characterization

Linear regression is an important step for obtaining the correct range of radius $r \in [r_{min}, r_{max}]$ that is needed to calculate the generalized fractal dimensions (defined by Equations 8 and 9) and the mass exponents (defined by Equation 4). We found an appropriate range of radii $r$ within the range of the interval located between 2 and 29 for linear regression. Thus, we selected this linear fit scaling range to perform multifractal analysis. We set the range of $q$ values from -7 to 12.

We calculated $\tau(q)$ and the $D(q)$ curves using the sandbox algorithm by Liu *et al.* [33] and based upon the shapes obtained from the spectrum in Figure 3, it can be seen that the curves are non-linear, suggesting that the network is multifractal.

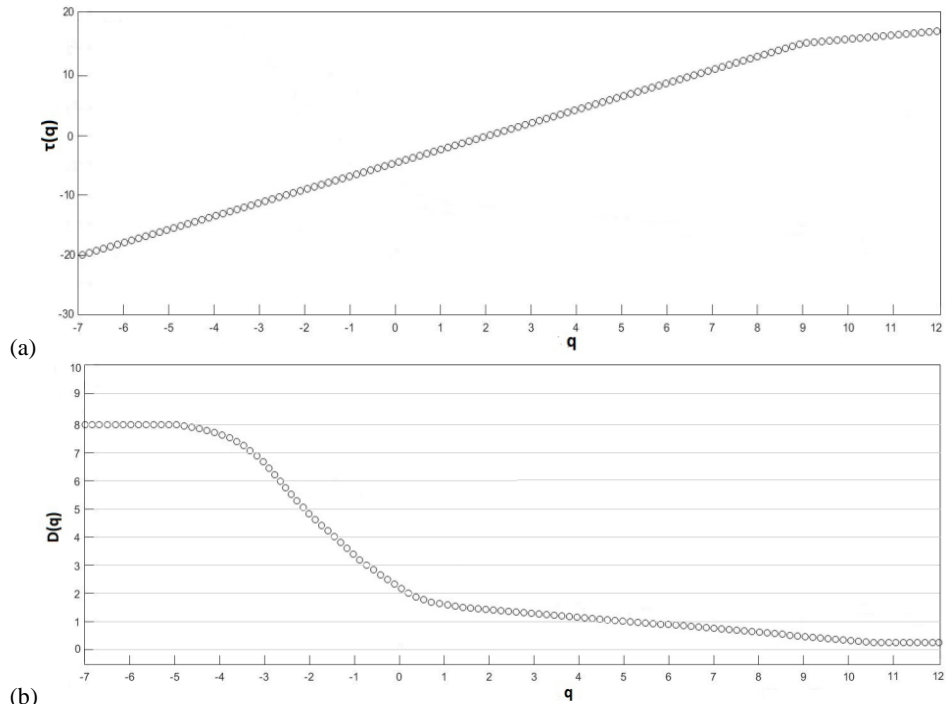

(a)

(b)

Figure 3 a) Plot of mass exponents $\tau(q)$ as function of $q$. b) Plot of generalized fractal dimensions $D(q)$ as function of $q$. Curves indicated by circles represent numerical estimations of the mass exponents and generalized fractal dimensions, respectively.

In Figure 3b, the $D(q)$ function decreases sharply after the peak reaches its end when $q$ is close to -4.5. This trait could be interpreted as high densities around the hubs in the network. The hubs have a high number of links connected to them; therefore the density of links around the sections near the hubs is higher than in other parts of the network. These hub nodes or important companies have a noticeable larger amount of business partners (for example: their own customers, or their suppliers or any other business parties that interact financially with them) than the rest of the companies in the network have, and it is interesting to observe that this characteristic can be explored and identified by looking at the values of $D(q)$ spectra. The multifractality seen in our network reveals that the system cannot be described by a single fractal dimension, suggesting that the multifractal approach provides a better characterization than the fractal approach; hence, this means that the Estonian economy is multifractal.

The quantity $\Delta D(q)$ describes density changes of links in our network. We use $\Delta D(q) = D(q)_{max} - \lim D(q)$ to observe how the values of $D(q)$ change along the spectrum. From Figure 3b, we found that $\lim D(q) = 0.37$ and $D(q)_{max} = 7.8$ and this means that $\Delta D(q) = 7.43$. A large value of $D(q)$ means that the distribution of links is very irregular in our network, suggesting that there are areas of hubs where the links are very densely grouped contrasting with areas where the nodes are connected with only just few links. For the economy of a country, this makes sense because not all the companies have the role of hubs in the network of payments; many companies are just small participants.

Table 3
Comparison of the maximum values of *D(q)* in different networks

| Network | Number of nodes | Highest $D(q)$ | Reference |
|---|---|---|---|
| **Pure fractal network** | **6222** | **2.8** | **[32]** |
| **Small world network** | **6222** | **6.6** | **[32]** |
| **Semi fractal network** | **6222** | **3.1** | **[32]** |
| **Sierpinski weighted fractal network** | **9841** | **2.0** | **[2]** |
| **Cantor dust weighted fractal network** | **9841** | **3.2** | **[2]** |
| **High-energy theory collaboration weighted network** | **8361** | **6.0** | **[2]** |
| **Astrophysics collaboration weighted network** | **16706** | **6.2** | **[2]** |
| **Computational geometry collaboration weighted network** | **7343** | **5.1** | **[2]** |
| **Barabási & Albert model scale-free network** | **10000** | **3.6** | **[33]** |
| **Newman and Watts model small-world network** | **10000** | **4.8** | **[33]** |
| **Erdös-Rényi random graph model** | **10000** | **3.9** | **[33]** |
| **Barabási & Albert model scale-free network** | **7000** | **3.4** | **[35]** |
| **Random network** | **5620** | **3.5** | **[35]** |
| **Random network** | **449** | **2.4** | **[35]** |
| **Protein-Protein interaction network: Human** | **8934** | **4.9** | **[35]** |
| **Protein-Protein interaction network: Arabidopsis thaliana** | **1298** | **2.5** | **[35]** |
| **Protein-Protein interaction network: C. elegans** | **3343** | **4.5** | **[35]** |
| **Protein-Protein interaction network: E. coli** | **2516** | **4.1** | **[35]** |
| **Small world network** | **5000** | **3.0** | **[35]** |
| **Estonian network of payments** | **16613** | **7.8** | **[45]** |

## 4. Conclusions

We presented the first multifractal analysis of a complex network of payments. We studied specific fractal and multifractal properties of a novel and unique network: the Estonian network of payments. In this study, we presented a fractal scaling analysis where we identified the underlying skeleton structure of the network. We calculated its fractal dimension and compared it with the fractal dimension of the original network. We found that the skeleton network had a slightly smaller fractal dimension than the original network. This comparison, between the fractal scaling in our original network and the corresponding skeleton network reveals that there are only slightly distinct patterns according to the fractality in the network. This means that the skeleton network preserves the structure very well while simplifying the complexity of the network. Then, the skeleton network captures the general structure of the network and by observing the properties of the skeleton, an easier visualization of the topological organization of the network can be achieved.

Fractal analysis helps to calculate and understand the fractal dimension of complex networks. However, it is necessary to describe and characterize the multiple fractal patterns which cannot be described by a single fractal dimension, thus we also performed a multifractal analysis to our network. Multifractal analysis allows the calculation of a set of fractal dimensions, particularly the generalized fractal dimensions. We examined the general multifractal structure and explored some statistical features of our network. In order to study the multifractal structure, we calculated the spectrum of the mass exponents $\tau(q)$ and the generalized fractal dimensions $D(q)$ curves, using a sandbox algorithm for multifractal analysis of complex networks adapted by Liu *et al.* [33]. This algorithm is based on the fixed-size box-counting algorithm developed by Tél *et al.* [46]. The sandbox algorithm utilized in this study could also be used to explore and characterize other similar kinds of economic networks.

Our results indicated that multifractality exists in the Estonian network of payments, and this suggests that the Estonian economy is multifractal (from the point of view of networks). We found large values of $D(q)$ spectra and this means that the distribution of links is quite irregular in the network, suggesting there are specific nodes which hold densely connected links, meanwhile other nodes hold just few links. This type of structure could be relevant when specific critical events occur in the economy that could threaten the whole network. It is important to continue observing, describing and analyzing the structures and characteristics of economic complex networks in order to be able to understand their underlying processes or to be able to detect patterns that could be useful for predicting or forecasting events and trends. The addition of evidence through empirical studies in favor of fractality and multifractality of economic networks represents a step forward towards the knowledge on the universality and the unraveling of the complexity of economic systems.

Further applications and studies could extend this topic by examining the potential factors that drive the strength of the multifractal spectrum. Some applications could involve studying the origin of such factors. Another interesting line of research would be to study the patterns and the changes of the multifractal spectrum across different periods of time. Particularly, it would be interesting to analyze such patterns during financial crisis periods for risk pattern recognition purposes. Also, it would be interesting to take into account different probability measures for such kind of multifractal analysis. Other direction of the studies could focus on building network models that attempt to forecast country money flows or potential industry growth trends based on transactions data.

## 5. Acknowledgements

We thank Swedbank (Estonia) for allowing us to retrieve the data set for the analysis. This research was supported by the European Union through the European Regional Development Fund (Project TK 124).